\documentclass[3p,12pt,authoryear]{elsarticle}

\usepackage{makecell}
\usepackage{booktabs}
\usepackage{amsmath}
\usepackage{amssymb}
\usepackage{graphicx}
\usepackage{hyperref}
\usepackage[toc,page]{appendix}

\makeatletter
\renewcommand\hyper@natlinkbreak[2]{#1}
\makeatother

\journal{Computers and Education: Artificial Intelligence}

\begin{document}

\begin{frontmatter}

\title{The Theoretical Foundation of Socratic Tests: \\ Dynamic, Multimodal, Conversational Examinations}

\author[1]{Ilya Mikhelson\corref{cor1}}
\ead{ilya@northwestern.edu}
\cortext[cor1]{Corresponding author}

\affiliation[1]{organization={Department of Electrical and Computer Engineering, Northwestern University}, city={Evanston}, state={IL}, country={USA}}

\begin{abstract}
Traditional static assessments rely on a subtractive, deficit-based grading model that often penalizes ambition and obscures diagnostic feedback. Conversely, traditional face-to-face oral examinations introduce severe construct-irrelevant variance by exacerbating performative anxiety and the sociological power imbalances inherent to academic hierarchies. This paper presents the theoretical foundation for the ``Socratic Test,'' an automated, computer-mediated conversational assessment. By integrating Dynamic Assessment principles, multimodal workspaces, Bloom's Taxonomy for real-time proctoring, and the SOLO Taxonomy for structural evaluation, the Socratic Test actively maps a student's cognitive boundaries. This paper formalizes the use of graduated scaffolding to quantify the Zone of Proximal Development (ZPD) and details a non-compensatory, additive grading architecture that prioritizes mastery over penalty and human-AI alignment to ensure unprecedented measurement reliability.
\end{abstract}


\end{frontmatter}

\section{Background and Motivation}

\subsection{Written Examinations - Reliability over Validity}
The prevailing assessment paradigm in higher education relies heavily on static, written examinations evaluated via a subtractive grading model \citep{feldman2023grading}. Students begin with a theoretical perfect score, and points are deducted for errors, omissions, or misapplications. While computationally efficient for instructors and highly \textit{reliable} (capable of being graded consistently), written exams often lack \textit{validity}. They frequently fail to measure true competence, allowing students to mask knowledge gaps through test-taking strategies or rote memorization \citep{frederiksen1984real, struyven2005students}. 

Furthermore, open-ended static assessments frequently suffer from the ``punishing ambition'' problem. While individual instructors may attempt to grade holistically, the underlying mathematical structure of a static exam systemically incentivizes risk aversion. For instance, a student who attempts a highly sophisticated, novel synthesis but makes a minor structural error is mathematically penalized compared to a student who safely executes a rudimentary, surface-level response. This subtractive grading model conflates diagnostic feedback with punitive measures, ultimately narrowing thinking, reducing intrinsic motivation, and conditioning students into compliance rather than deep engagement \citep{feldman2023grading, kohn1993punished}.

\subsection{Oral Examinations - The Standardization Fallacy}
While there are numerous alternative to written examinations, such as essays, portfolios, and performances, the most direct alternative for probing a student's cognitive boundaries has been the oral examination. Oral exams exhibit incredibly high validity, as they probe the edges of a student's domain knowledge and professional readiness \citep{huxham2012oral, joughin1998dimensions}. However, their traditional reputation is complicated by two major critiques: one a genuine limitation, and the other a pedagogical illusion.

The genuine limitation is the introduction of significant construct-irrelevant variance. Face-to-face interrogations raise a student's Affective Filter, a psychological barrier of anxiety and fear of judgment that blocks working memory \citep{krashen1982principles}. Consequently, oral exams often measure a student's public speaking proficiency and stress management rather than their pure domain knowledge, with literature frequently citing the format as a source of severe, debilitating anxiety for undergraduate students \citep{joughin1998dimensions, iannone2020impact, laurin2016oral}.

The second critique is a perceived lack of equity. Because a proctor must ask different questions to different students to map their unique knowledge bounds, critics argue the exam is not standardized. However, this critique is an illusion stemming from the Standardization Fallacy, i.e., the misconception that standardizing the student's \textit{experience} (identical questions) is required to standardize the \textit{measurement} \citep{wainer2000computerized, lord2012applications}. The theoretical precedent for dismissing this fallacy is firmly established in Computerized Adaptive Testing (CAT) \citep{lord2012applications}. High-stakes professional licensure exams (e.g., NCLEX, USMLE) and admissions exams (e.g., GMAT, SAT) operate on adaptive algorithms where no two candidates receive the same questions. True equity is achieved through criterion-referenced grading, where the specific conversational path varies, but the structural threshold required to prove mastery remains immutable \citep{biggs1996enhancing}.

\section{Theoretical Framework of the Socratic Test} \label{sec:framework}
To resolve the tension between the reliability of written exams and the validity of oral exams, I propose the Socratic Test. This framework utilizes an adaptive, typed conversational assessment driven by artificial intelligence (AI) to standardize the methodology of measurement without standardizing the specific conversational prompts. To achieve this, the platform relies on two foundational pedagogical frameworks: one to govern the complexity of the proctor's questions, and another to measure the structural quality of the student's answers. These frameworks are defined, compared, and contrasted in Table \ref{tab:taxonomies}.

\begin{table}[htbp]
    \centering
    \caption{Foundational Cognitive Frameworks of the Socratic Test}
    \label{tab:taxonomies}
    \begin{tabular}{p{1cm} p{5cm} p{5cm}}
        \toprule
        \textbf{Level} & \textbf{Bloom's Taxonomy (Prompt Objective)} \citep{krathwohl2002revision} & \textbf{SOLO Taxonomy (Response Structure)} \citep{biggs2014evaluating} \\
        \midrule
        \textbf{1} & \textbf{Remember:} Retrieve relevant factual knowledge and definitions from long-term memory. & \textbf{Prestructural:} The response misses the point entirely or relies on irrelevant information. \\
        \addlinespace
        \textbf{2} & \textbf{Understand:} Construct meaning from instructional messages (e.g., interpreting, summarizing). & \textbf{Unistructural:} The response correctly identifies or utilizes a single relevant aspect. \\
        \addlinespace
        \textbf{3} & \textbf{Apply:} Carry out or use a procedure in a specific, given situation. & \textbf{Multistructural:} The response identifies several relevant independent aspects but fails to connect them. \\
        \addlinespace
        \textbf{4} & \textbf{Analyze:} Break material into constituent parts and determine how they relate to one another. & \textbf{Relational:} The response successfully integrates the disparate aspects into a coherent structure. \\
        \addlinespace
        \textbf{5} & \textbf{Evaluate:} Make structural judgments based on established criteria and standards. & \textbf{Extended Abstract:} The response generalizes the integrated principle to a completely novel domain. \\
        \addlinespace
        \textbf{6} & \textbf{Create:} Reorganize elements into a novel, coherent whole or original product. & \textit{(The SOLO framework concludes at Level 5)} \\
        \bottomrule
    \end{tabular}
\end{table}

\subsection{Dynamic Assessment and the ZPD}
The Socratic Test is rooted in Dynamic Assessment (DA) \citep{lantolf2004dynamic}, derived from Lev Vygotsky's theory of cognitive development \citep{vygotsky1978development}. Unlike static assessments, which only measure what a student has already mastered, DA integrates instruction and assessment to measure a student's Zone of Proximal Development (ZPD), i.e., the space between what a learner can do independently and what they can achieve with guidance \citep{lantolf2004dynamic}. By observing how a student responds to conversational hints, the system mathematically differentiates between \textit{Independent Performance} (unassisted success) and \textit{Assisted Performance} (success achieved via scaffolding). Evaluating the full breadth of the ZPD yields a much truer measure of systemic understanding, as it prevents a student's demonstrated competence from being artificially capped by a momentary lapse in foundational memory.

\subsection{Multimodal Evidence and Cognitive Offloading}
A pure text-based chat is often insufficient for evaluating highly complex, spatial, or mathematical competencies across various disciplines. The Socratic Test interface includes integrated workspaces, such as a whiteboard, an integrated development environment (for code), a calculator, and a long-form response interface, which serve a critical pedagogical function known as Cognitive Offloading \citep{kirsh1995intelligent}.

By allowing students to sketch diagrams or draft code, the workspaces act as external working memory, reducing intrinsic cognitive load \citep{sweller1988cognitive}. More importantly, capturing this offloaded work natively within a digital drawing and text interface provides the AI with a significantly richer, multimodal evidence trail. This interface specifically protects students who may struggle with English prose articulation; rather than relying solely on language proficiency, the AI evaluates the structural logic of a hand-drawn schematic or a mathematical derivation. Furthermore, transitioning from a face-to-face interrogation to a typed, asynchronous interface is highly likely to lower the student's affective filter. Decades of research into Computer-Mediated Communication (CMC) demonstrate that typed interfaces significantly reduce communication apprehension and performative anxiety \citep{satar2008effects, high2009social, krashen1982principles}. By removing the human authority figure, the interface also mitigates the intimidating power dynamics of traditional oral testing, creating an environment focused on higher-order cognitive skills rather than stress management.

\subsection{Proctoring with Bloom's Taxonomy} \label{sec:proctoring}
To navigate the ZPD systematically, the AI proctor utilizes Bloom's (revised) Taxonomy \citep{krathwohl2002revision} to structure its prompting. The proctor begins with low-cognitive-load questions and incrementally elevates the complexity. While cognitive mapping is not strictly linear, Bloom's Taxonomy provides the AI with an efficient, structured heuristic for navigating the ZPD. Foundational recall (rote memory) serves as the indispensable vocabulary of any discipline \citep{willingham2009why}. Establishing this vocabulary early allows the proctor to verify that the student possesses the necessary foundational tools before advancing to complex analysis. When a student struggles, the proctor employs the System of Least Prompts (graduated prompting) \citep{campione1987linking}, offering a strict hierarchy of scaffolding:
\begin{enumerate}
    \item General Nudge: A non-directive prompt to rethink the premise.
    \item Specific Cue: A directive prompt pointing to a specific missing concept.
    \item Targeted Scaffold: A highly constrained prompt isolating the exact point of failure.\item Direct Instruction (Graceful Exit and Pivot): If the student exhausts the prior three hints, the AI explicitly provides the missing foundational knowledge (e.g., giving the student the forgotten formula). Crucially, this does not terminate the topic. Because knowledge is not strictly hierarchical, a student may fail a Bloom 1 recall question but still possess Bloom 4 relational mastery \citep{krathwohl2002revision, agarwal2019retrieval}. By supplying the missing definition, the AI bridges the gap, allowing the conversation to proceed.
\end{enumerate}
Therefore, exhausting the scaffolding hierarchy on a single prompt does not prematurely terminate the entire topic, nor does it automatically finalize the current tier. Instead, it finalizes the student's score for that specific interaction, effectively defining their Cognitive Ceiling for that discrete skill. The AI logs the interaction, awards zero points to the student's earned score, and utilizes the Graceful Exit to explicitly provide the missing foundational knowledge. 

Rather than arbitrarily unlocking the next cognitive tier, this failed interaction is mathematically accounted for via a ``Shadow Ledger'' (as detailed in Section \ref{sec:state_machine}). The proctor will continue generating parallel prompts within the current cognitive tier until the total evidence buffer is mathematically satisfied. The AI only pivots vertically to the next tier, or horizontally to an entirely new competency, once these rigorous evidentiary thresholds have been met, ensuring the student is never artificially capped by a single momentary lapse in recall.

\subsection{Assessing with the SOLO Taxonomy}
While Bloom's Taxonomy dictates the cognitive difficulty of the \textit{prompt}, the AI utilizes the SOLO Taxonomy (Structure of the Observed Learning Outcome) to measure the structural quality of the \textit{response} \citep{biggs2014evaluating}.

The decision to decouple the prompting framework from the evaluation framework is rooted in the pursuit of grading objectivity and transparency. In traditional static assessments, or even when attempting to evaluate responses using Bloom's Taxonomy, grading is often inherently subjective. A student reviewing a marked exam might argue that their answer demonstrated ``Analysis'' rather than mere ``Application,'' leading to contentious appeals over partial credit. The SOLO Taxonomy neutralizes this ambiguity by shifting the evaluation from semantic interpretation to structural complexity. The grading criteria become strictly observable and defensible: a student's response objectively contains either a single isolated fact (Unistructural), multiple unconnected facts (Multistructural), or logically integrated facts (Relational), as detailed in Table \ref{tab:taxonomies}. This structural clarity makes the rubric highly transparent, streamlines the post-exam appeals process (Section \ref{sec:grading}), and fosters deep student trust in the platform's fairness.

A natural pedagogical question arises: how does the AI anchor its application of SOLO to a highly specific, advanced topic during the live conversation, before the instructor has provided any graded examples? During real-time proctoring, the AI relies on a discipline-agnostic, zero-shot system prompt anchored strictly to the universal structural definitions of the taxonomy. Rather than attempting to evaluate domain-specific semantic nuance in real-time, the AI acts as a structural parser.

This tentative, real-time assessment acts purely as a navigational routing mechanism to keep the conversation flowing. The system operates conservatively; if the AI is uncertain whether a response meets the expected structural threshold, it explicitly asks the student for clarification rather than advancing. Crucially, any inaccuracies in this baseline real-time anchoring are mathematically absorbed by the Oversampling Evidence Buffer (detailed in Section \ref{sec:buffer}) and subsequently corrected during the deterministic, post-exam grading pipeline (Section \ref{sec:grading}).

\section{Test Progression and Exam Configuration} \label{sec:progression}
\subsection{Mastery Accumulation and Dual Proctoring Modes}
In the Socratic Test's additive grading model, students start at zero and accumulate evidence of knowledge. However, raw accumulation introduces the gamification loophole of ``grinding'', where a student answers dozens of foundational questions to achieve a high score without ever demonstrating higher-order thinking.  To prevent this, the architecture rejects Compensatory Grading (where low-level skills can mathematically compensate for a lack of high-level skills) in favor of Non-Compensatory Grading. Rooted in the theory of Constructive Alignment \citep{biggs1996enhancing} (i.e., the pedagogical principle that assessment tasks and grading criteria must strictly align with intended learning outcomes), the system utilizes ``capped buckets'' for different cognitive tiers, configured by the instructor prior to the exam via one of two distinct proctoring modalities: Stair-Step Mode (Section \ref{sec:stair}) or Organic Mode (Section \ref{sec:organic}).

\subsubsection{Stair-Step Mode - Structured Scaffolding} \label{sec:stair}
In Stair-Step mode, the instructor defines a rigorous 2D grading matrix while setting up the exam. They configure distinct cognitive tiers (e.g., Foundational, Application, Synthesis) and map specific Bloom's levels to each tier. For example, an introductory course may map Bloom 1 and 2 to Foundational, while an advanced seminar might eliminate Bloom 1 entirely and begin the Foundational tier at Bloom 3. 

The instructor then defines the overarching exam topics. For each topic, they assign a specific point capacity to each cognitive tier, as well as a target time limit. Each cell in this matrix acts as a capped bucket. An example of such a matrix for an introductory Economics course can be seen in Table \ref{tab:matrix_concept}. Because the buckets are topic- and tier-specific, a student cannot mathematically compensate for dodging an Application question on Market Failure by answering an excess of Foundational questions on Supply \& Demand.

\begin{table}[tbph]
	\caption{Conceptual 2D Non-Compensatory Grading Matrix (Stair-Step Mode)}
	\label{tab:matrix_concept}
	\hskip-0.6cm
	\begin{tabular}{l c c c | c}
		\toprule
		\textbf{Topic ($t \in T$)} & \textbf{Foundational} & \textbf{Application} & \textbf{Synthesis} & \textbf{Topic Max} \\
		\midrule
		Topic A (e.g., Supply \& Demand) & Max 10 & Max 10 & Max 10 & \textbf{30} \\
		Topic B (e.g., Market Failure) & Max 10 & Max 15 & Max 15 & \textbf{40} \\
		Topic C (e.g., Elasticity) & Max 10 & Max 10 & Max 10 & \textbf{30} \\
		\midrule
		\textbf{Cognitive Tier Max} & \textbf{30} & \textbf{35} & \textbf{35} & \textbf{Total: 100} \\
		\bottomrule
	\end{tabular}
\end{table}

\subsubsection{Organic Mode - Fluid Exploration} \label{sec:organic}
While Stair-Step mode is ideal for structured knowledge assessment, Organic mode is designed for fluid, holistic exploration, akin to a traditional graduate defense. In Organic mode, the 2D matrix collapses into a 1D vector. The instructor no longer defines rigid cognitive tiers; instead, they define the broad exploration space by selecting the allowable Bloom's levels for the exam (e.g., testing exclusively at Bloom 3 through 6). 

Correspondingly, instructors assign point capacities broadly at the topic level, rather than the tier level. The proctor initiates an open-ended conversational anchor and allows the interaction to evolve organically, shifting Bloom's levels in real-time based on the student's conversational direction until the topic's overall point cap is satisfied. Because every individual interaction is still tagged with a specific Bloom's level and hint count, the underlying mathematical grading engine remains entirely unchanged.

\subsection{Mathematical Formulation}
To formalize this non-compensatory structure, the core system variables and their domains are defined in Table \ref{tab:variables}.

\begin{table}[htbp]
	\centering
	\caption{System Variables and Nomenclature}
	\label{tab:variables}
	\begin{tabular}{l p{8cm} l}
		\toprule
		\textbf{Variable} & \textbf{Definition} & \textbf{Domain / Range} \\
		\midrule
		$t$ & A specific domain topic being tested. & $t \in T$ \\
		$l$ & A cognitive grading tier (e.g., Foundational). & $l \in L$ \\
		$b$ & The Bloom's Taxonomy level of the prompt. & $b \in \{1,2,3,4,5,6\}$ \\
		$s$ & The SOLO Taxonomy level of the response. & $s \in \{1,2,3,4,5\}$ \\
		$k$ & The number of scaffolding prompts required. & $k \in \{0, 1, 2, 3, 4\}$ \\
		$\gamma_k$ & The scaffolding discount factor. & $\gamma_k \in [0, 1]$ \\
		$V(b, s)$ & Base point value (monotonically increasing with $b$). & $V \ge 0$ \\
		$P(b, s, k)$ & Total points generated by a single interaction. & $P \ge 0$ \\
		$M_{t,l}$ & Maximum allowable points (cap) for a specific bucket. & $M_{t,l} > 0$ \\
		\bottomrule
	\end{tabular}
\end{table}

Let $t \in T$ be a specific topic and $l \in L$ be a cognitive tier (e.g., Foundational, Application). Let $b \in \{1,2,3,4,5,6\}$ be the Bloom's level of the prompt. The tiers represent a mapping of multiple Bloom's levels.  For instance, Bloom 1 and 2 can map to the Foundational tier. Let $s \in \{1,2,3,4,5\}$ be the SOLO level of the response, and $k \in \{0, 1, 2, 3\}$ be the number of scaffolding prompts required. The point value $P$ generated by a single interaction is:

\begin{equation}
P(b, s, k) = \gamma_k V(b, s)
\label{eq:P}
\end{equation}

where $V(b, s)$ is the base point value of achieving SOLO level $s$ on a Bloom level $b$ prompt, and $\gamma_k \in [0, 1]$ is the scaffolding discount factor, quantifying the reduction in value from Independent to Assisted Performance. For example, an instructor may configure $\gamma_0 = 1.0$ (no hints, full credit), $\gamma_1 = 0.8$ (one hint), down to $\gamma_4 = 0.0$ for a Graceful Exit.  Crucially, $V(b, s)$ is a monotonically increasing function with respect to $b$; a simple Recall prompt (Bloom 1) yields fewer points than a complex Create prompt (Bloom 6).

To enforce the non-compensatory structure, the total grade is calculated by aggregating points across all specific buckets (topic $t$, tier $l$):

\begin{equation}
\text{Total Score} = \sum_{t \in T} \sum_{l \in L} \min \left( M_{t,l}, \sum P_{t,l} \right)
\label{eq:score}
\end{equation}

where $M_{t,l}$ represents the maximum allowable points for that specific matrix cell, and $P_{t,l}$ is the point value from Eq. \eqref{eq:P} earned within that cell. Once a bucket is full, the AI forces vertical or horizontal progression; answering further low-level questions within that topic yields zero additional points toward the final grade.

\textbf{A complete simulated transcript demonstrating the test progression of Sections \ref{sec:framework} and \ref{sec:progression} is provided in Appendix \ref{app:example}.}

\section{System Architecture and Implementation}
To deploy this theoretical framework safely within high-stakes academic environments, I developed \href{https://socratictest.com}{socratictest.com}, a custom software platform explicitly engineered to administer the Socratic Test modality. While recent literature has explored the use of Large Language Models (LLMs) as conversational tutors for formative feedback \citep{favero2024enhancing, liu2024socraticlm}, \href{https://socratictest.com}{socratictest.com} is specifically architected as a summative assessment engine. The platform relies on a decoupled architecture separating the real-time proctoring engine from the final deterministic gradebook.

\subsection{Proctoring Mechanics - Buffers, Ledgers, and Pivoting} \label{sec:state_machine}
During the live exam, the AI navigates an internal state machine governed by continuous mathematical accounting. The proctor's primary imperative is to gather sufficient conversational evidence to satisfy the point caps defined in the instructor's grading matrix (Table \ref{tab:matrix_concept}). To execute this equitably, the state machine utilizes three core mechanics, detailed below.

\subsubsection{The Evidence Buffer and Oversampling} \label{sec:buffer}
Because the AI's real-time assessment of a student's SOLO level is strictly provisional (awaiting instructor calibration in Step 2 of the grading pipeline (Section \ref{sec:grading})), the state machine cannot rely on strict minimum SOLO thresholds to gate student progression. Doing so risks trapping a student behind an AI hallucination. 

Instead, the system utilizes an Evidence Buffer parameterized by an Oversampling Factor. Prior to the exam, the instructor defines an oversample rate (e.g., 120\%). During a Stair-Step exam, if a tier requires 10 points to fill, the AI will continue prompting the student until its provisional real-time evaluation assesses that 12 points worth of evidence have been gathered. This mathematical buffer protects the student; if the instructor retroactively downgrades a specific interaction's SOLO score during post-exam calibration, the oversampled evidence buffer ensures the student was not unfairly denied the opportunity to earn the tier's maximum points.

\subsubsection{The Shadow Ledger}
To maintain exam integrity, the system must prevent students from exploiting the conversational interface by skipping difficult questions or intentionally exhausting hints to bypass a topic. The platform achieves this via a hidden Shadow Ledger of ``attempted points.''

When a student explicitly requests to skip a prompt, or when they exhaust the scaffolding hierarchy (more than 3 hints, detailed in Section \ref{sec:proctoring}) triggering a Graceful Exit, the AI complies and pivots. However, it simultaneously logs the baseline point value of the abandoned prompt into the Shadow Ledger. This baseline value is deterministically anchored to the Expected Strong Response defined in Table \ref{tab:shadow_ledger} (e.g., skipping a Bloom 3 prompt logs the expected $V(3,4)$ point value into the ledger). The Vertical Gate (i.e., the blocker to proceed to a higher tier) for a cognitive tier only opens when the sum of the student's \textit{Earned Points} plus their \textit{Shadow Ledger Points} meets the oversampled tier capacity. 

Consequently, skipping a question mathematically accelerates the closure of the cognitive tier without awarding the student evidence points, permanently limiting their potential score. Conversely, if a student struggles but eventually arrives at the correct answer using hints, no shadow points are levied; they simply earn their standard, fractionally discounted points, preserving their ability to continue gathering evidence in that tier.

\begin{table}[htbp]
    \centering
    \caption{Shadow Ledger Baseline Matrix: Anchoring Bloom's Prompts to Expected SOLO Responses}
    \label{tab:shadow_ledger}
    \begin{tabular}{p{4cm} p{4cm} p{4cm}}
        \toprule
        \textbf{AI Prompt \newline (Bloom's Level)} & \textbf{Prompt Ceiling \newline (Max SOLO)} & \textbf{Expected Strong Response Baseline (Shadow Ledger)} \\
        \midrule
        Bloom 1 & SOLO 3 & SOLO 2 \\
        Bloom 2 & SOLO 3 & SOLO 3 \\
        Bloom 3 & SOLO 4 & SOLO 4 \\
        Bloom 4 & SOLO 4 & SOLO 4 \\
        Bloom 5 & SOLO 5 & SOLO 4 \\
        Bloom 6 & SOLO 5 & SOLO 5 \\
        \bottomrule
    \end{tabular}
\end{table}

\subsubsection{User-Directed Pivoting}
To maximize student agency and cognitive offloading, the platform abandons forced temporal constraints in favor of user-directed pivoting. At any point, a student can access a navigation menu detailing the time spent on each topic and elect to pivot horizontally to a new domain. 

This pivot incurs no penalty. When the student eventually returns to the abandoned topic, the state machine resurrects the exact conversational context, Evidence Buffer, and Shadow Ledger state from the moment of departure, preventing pivoting from being used as an evasive tactic. (Note: Students may pivot between overarching topics, but they cannot manually pivot between cognitive tiers within a topic, as foundational tiers must be completed to scaffold advanced topics). Furthermore, when a student exceeds the instructor-defined target time for a topic, the interface issues a non-blocking visual warning. This timer grounds the student's pacing without forcefully interrupting their cognitive flow.

\subsection{Mitigating Hallucination and the ``Out-of-Scope'' Protocol} \label{sec:audit}
A primary concern when deploying LLMs in assessment is the risk of AI hallucination, where the proctor might ask questions about out-of-scope material or state incorrect premises. The Socratic Test handles this through an asynchronous audit mechanism that is strictly mathematically bound to the real-time Shadow Ledger to prevent gamification.

During the live exam, if a student claims a premise is out-of-scope or identifies an AI error, the proctor accepts the student's assertion, flags the interaction for audit, and pivots to a new question.  However, the state machine must prevent the student from exploiting this feature to infinitely cycle questions. Therefore, the system initially treats an ``Out-of-Scope'' flag identically to a standard ``Skip.'' It immediately logs 100\% of the prompt's baseline point value (Table \ref{tab:shadow_ledger}) into the student's Shadow Ledger, and the AI pivots to the next prompt. This provides a critical real-time safeguard: a student continuously claiming out-of-scope will mathematically exhaust the tier's Evidence Buffer, definitively limiting their potential score and moving the exam forward.

In a traditional subtractive system, adjudicating a dodge requires applying an arbitrary penalty. In the Socratic Test's additive, non-compensatory framework, no punitive deduction is necessary. If a student maliciously dodges a valid topic, they simply fail to fill the non-compensatory bucket for that specific domain, inherently limiting their final grade.

The equity of this mechanism relies entirely on the post-exam instructor audit via a four-point scale:

\begin{enumerate}
    \item Student Correct: The instructor verifies the AI erred. Because identifying a structural error proves advanced domain mastery \citep{krathwohl2002revision}, the system converts the initial Shadow Ledger penalty directly into Earned Points for that tier. The student gets full credit for the skipped prompt.
    \item Valid Confusion: The premise was technically valid, but poorly phrased. The system gives the student the benefit of the doubt, converting the Shadow points into Earned points.
    \item Partial Evasion: The student deflected a valid premise but attempted some engagement. The instructor retroactively flags the interaction as a partial skip. The system leaves 50\% of the point value in the Shadow Ledger as a permanent penalty, converting the remaining 50\% to Earned points.
    \item Complete Evasion: The student abused the mechanism to dodge a valid question, or they did not know how to answer the question.  In either case, it is identical to a Skip. The 100\% Shadow Ledger penalty applied during the live exam remains permanent.
\end{enumerate}

By shifting the burden of trust from the live AI to the post-exam audit, this architecture completely neutralizes the risk of students gaming the conversational interface, while guaranteeing they are mathematically rewarded for catching an algorithmic mistake.  An example of this can be seen in Appendix \ref{app:example}.

\subsection{Interaction-Level Grading and Human-AI Alignment}
A persistent critique of integrating AI into higher education is the perceived loss of instructor oversight and the inherent unreliability of algorithmic evaluation. However, this skepticism often ignores the profound fallibility of traditional human grading. Literature confirms that human marking suffers from severe inter-rater and intra-rater reliability issues, driven by grader fatigue and fluctuating internal standards \citep{bloxham2011mark}. 

The Socratic Test platform mitigates the unreliability of both humans and autonomous algorithms by deploying a deterministic, interaction-level grading pipeline. Rather than attempting to grade an entire transcript holistically, which introduces systemic variance, the architecture isolates the single subjective variable (the SOLO level) from the objective variables recorded during the live exam (the Bloom's level and the hint count). To ensure absolute fidelity, the system utilizes a calibration and validation loop as part of a 7-step grading pipeline, detailed below.

\subsubsection{The 7-Step Alignment and Grading Pipeline} \label{sec:grading}

\textbf{Step 1: The Out-of-Scope Audit and Global Rules.} 
During the live assessment, the proctor automatically tags interactions where a student flags a premise as out-of-scope or identifies an AI hallucination. The instructor rapidly audits these real-time tags. If a hallucination is verified, the flagging student is rewarded (as detailed in Section \ref{sec:audit}). Crucially, to protect systemic equity, this verification automatically generates a ``Global Exclusion Rule'' within the grading engine. This ensures that any other student in the cohort who encountered the exact same hallucinated premise, but lacked the assertiveness to challenge the AI, is mathematically protected.

\textbf{Step 2: Ambiguity Calibration.} 
Following the audit, the AI performs an initial sweep of every interaction across the entire cohort. It predicts a provisional SOLO score for each interaction and calculates a confidence interval. The system then isolates the 20 most ambiguous interactions for calibration. To prevent a single highly unconventional student from skewing the model, the system enforces a strict cap of 4 calibration interactions per student. The instructor manually reviews these 20 edge-case interactions and assigns the definitive SOLO scores, establishing the semantic baseline.

\textbf{Step 3: Inter-Rater Validation.} 
Using the calibrated examples, the AI attempts to assign SOLO scores to a new, random validation set of 15 interactions. The instructor reviews these interactions and simply marks ``Agree'' or ``Disagree'' with the AI's assessment. If the instructor disagrees with 2 or more interactions, those failed interactions are fed back into the calibration pool (Step 2), and the loop is repeated. The system only unlocks the mass grading phase once strong algorithmic alignment is mathematically proven (i.e., at most 1 disagreement during validation).

\textbf{Step 4: Contextual Mass Grading.} 
Once validated, the AI utilizes the calibration data from Step 2 and the Global Exclusion Rules from Step 1 to mass-grade the entire cohort. While the grading math is executed discretely at the interaction level, the AI is fed the entirety of each student's transcript to ensure it possesses the full conversational context when evaluating an individual response. Because the Bloom's level ($b$) and the scaffolding discount factor ($k$) were recorded deterministically during the live exam, the AI's assignment of the SOLO score ($s$) allows the system to instantly and deterministically calculate the point value of every interaction using Eq. \eqref{eq:P}.

\textbf{Step 5: Instructor Review and Override.} 
Following mass grading, the complete transcripts and their deterministic scores are presented to the instructor, alongside any specific interactions the AI flagged as highly unusual during Step 4. The instructor adjudicates these flags and reviews the transcripts. If the instructor disagrees with an AI-assigned SOLO score for any reason, they can manually override it. Crucially, to maintain systemic equity, any overridden interaction is automatically converted into a new calibration example, and the entire cohort is seamlessly regraded. This ensures that any ad-hoc grading leniency or strictness is applied universally to all students.

\textbf{Step 6: Publication and Curve.} 
Once the instructor is fully satisfied with the transcript reviews, an optional statistical curve can be applied to the deterministic totals. The final grades and the fully marked-up transcripts are then published and made visible to the students on their dashboards.

\textbf{Step 7: The Appeals Queue.} 
Upon reviewing their marked-up transcripts, students have the option to appeal the specific SOLO grade of any individual interaction. To do so, the student must submit a written justification defending why their response warrants a higher structural classification. These appeals populate an internal queue for the instructor, who can review the specific interaction and choose to either uphold or overrule the grade, structurally closing the pedagogical feedback loop while ensuring total transparency.

\section{Student Experience}
\begin{figure}[htbp]
    \centering
    \includegraphics[width=0.9\textwidth]{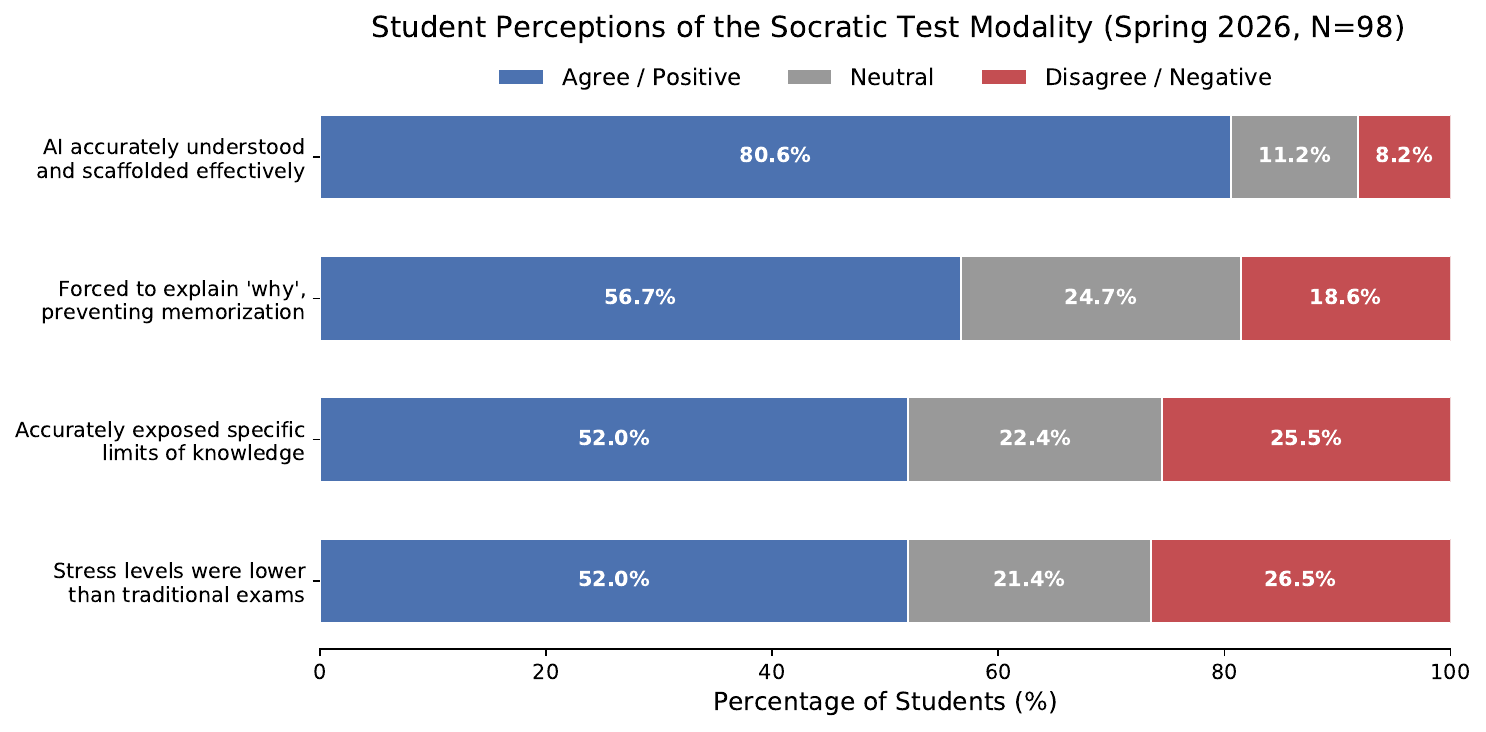}
    \caption{Student Perceptions of the Socratic Test Modality (Spring 2026, $N=98$)}
    \label{fig:student_survey}
\end{figure}

A critical barrier to faculty adoption of automated oral examinations is the assumption that students will be uniformly intimidated by interacting with an AI proctor. However, recent literature regarding generative AI in higher education, combined with empirical data from a Spring 2026 pilot deployment of the Socratic Test across three university courses (quantitative results in Figure \ref{fig:student_survey}), suggests a more nuanced reality. While a minority of students experienced friction, a significant majority were highly receptive to conversational AI, provided their concerns regarding fairness and accuracy were structurally addressed. (It should be noted that while these pilot results are promising, future controlled studies directly comparing AI-mediated exams against traditional face-to-face oral exams and static written exams are required to fully isolate the specific variables driving student acceptance.)

\subsection{Mitigating the Affective Filter} \label{sec:affective}
While traditional oral exams are notorious for raising a student's Affective Filter, they also disproportionately penalize students who are unaccustomed to the social dynamics of higher education. Face-to-face interrogations frequently test a student's mastery of the ``Hidden Curriculum'' (i.e., the unwritten rules, power dynamics, and cultural norms of academia), rather than their actual domain competence \citep{sellers2023message}.

The Socratic Test's asynchronous, typed interface demonstrably neutralizes this power imbalance. In the Spring 2026 pilot cohort, 52\% of students reported that their stress levels were lower during the AI assessment compared to traditional written exams, while an additional 21.4\% reported no change in stress.  By upending the traditional academic hierarchy and replacing an intimidating human evaluator with a neutral conversational partner, the platform provides students with the time for relaxation and the psychological safety required for deep cognitive synthesis.

Qualitative feedback highlighted the reduction in performative anxiety, with one student noting, \textit{``I think it was way less stressful and was a better assessment of my understanding.''} While a minority of students noted that the novelty of the format induced initial anxiety, they largely recognized its long-term pedagogical value: \textit{``I think the novelty of the testing format was stressful, but overall this will be a significantly better testing mechanism.''}

\subsection{Technology Acceptance and the Value of Scaffolding}
Research utilizing the Technology Acceptance Model (TAM) and the Unified Theory of Acceptance and Use of Technology (UTAUT) indicates that students' intention to use conversational AI is heavily driven by perceived usefulness and immediate, context-sensitive feedback \citep{strzelecki2024use}. Unlike a static exam, which is purely evaluative, students perceive conversational AI as a personalized tool that clarifies complex academic concepts while assessing them \citep{crompton2023artificial}.

The pilot data heavily supports this finding. Over 80\% of surveyed students agreed that the AI accurately understood their typed responses and pushed them appropriately when their answers were incomplete. Students explicitly praised the proctor's use of the System of Least Prompts (graduated scaffolding) to help them navigate their Zone of Proximal Development. As one student observed: \textit{``I couldn't remember the formula for moment of inertia, but it showed the units for moment so I would intuitively determine its formula.''} Another noted, \textit{``Instead of moving on, it explained more into detail what they were asking and then it helped me answer.''}

\subsection{Moving Beyond Recall}
By forcing students to articulate their logic, the modality successfully combats rote memorization. The majority of pilot participants agreed that the conversational format forced them to explain the \textit{``why''} behind their answers, preventing them from relying on surface-level recall. Furthermore, students recognized the system's ability to find their Cognitive Ceiling, with over half agreeing that the exam accurately exposed the specific limits of their knowledge. As one participant described the adaptive scaling: \textit{``It asked me smaller questions leading up to the main question until I understood.''}

\subsection{Critical Awareness and Trust}
Crucially, students are not blindly trusting these tools. Literature shows that perceived risk, specifically regarding AI hallucination, data privacy, and the accuracy of evaluation, is a significant deterrent to student adoption \citep{cotton2024chatting}. Students possess a high degree of critical awareness and fear being penalized for an AI proctor's error.

The Socratic Test explicitly neutralizes this perceived risk through its ``Out-of-Scope'' audit protocol. By granting students the explicit authority to challenge the AI's premises and forcing a pivot in questioning when a challenge is issued, the platform structurally empowers the learner. This mechanism perfectly aligns with student anxieties regarding generative AI; it builds deep trust in the assessment modality by guaranteeing that human oversight (the instructor's audit) remains the ultimate arbiter of algorithmic fairness.

\section{Faculty Experience}
\subsection{High-Resolution Differentiation}
A primary grievance with static examinations is their inability to evaluate multiple proficiency thresholds simultaneously. A static test is often calibrated to distinguish Pass/Fail or separate the highest achievers, but struggles to accurately differentiate the intermediate boundaries. 

The Socratic Test's adaptive state machine eliminates this limitation. As one pilot instructor in the engineering cohort noted, the AI \textit{``titrates its way to the limit of each student's understanding.''} Because the proctor scales the cognitive burden dynamically, the instructor was able to \textit{``directly compare transcripts to differentiate between students at the D/F boundary as well as the A/B boundary,''} while simultaneously presenting \textit{``multiple 'stretch' lines of questioning''} to the top quartile of the cohort.

\subsection{Eliminating Construct-Irrelevant Variance}
Static assessments in quantitative fields frequently suffer from construct-irrelevant variance by conflating conceptual domain mastery with computational speed. The pilot instructor noted that traditional exams \textit{``often award undue credit to students who are particularly strong in math/calculus/calculations and are quick and accurate with the 'grind'.''} 

By utilizing the Socratic Test, the instructor successfully decoupled the mathematical execution from the conceptual physics. Utilizing a mastery-based rubric, the instructor isolated ``Fluid Mechanics'' objectives from ``Math Competency,'' ensuring that the assessment strictly measured the intended construct rather than acting as a proxy test for calculus proficiency.

\subsection{Real-Time Disambiguation and Regrade Mitigation}
In traditional open-ended assessments, excellent students often lose points by going off on valid but unintended tangents, or by making alternative assumptions that clash with the static rubric. 

The conversational interface resolves this ambiguity in real-time. The pilot instructor highlighted a specific interaction where a student defined their coordinate system based on a diagram arrow, while the AI proctor initially assumed a standard ``right is positive'' framework. In a static exam, this misalignment would result in a heavily penalized answer and a subsequent regrade request. However, the AI engaged in a \textit{``very short exchange [that] resolved this difference of opinion.''} The student was able to seamlessly defend their premise, proving deep conceptual mastery and transforming a potential point of friction into what the student described as a \textit{``9-out-of-10 testing experience.''}

\subsection{Transparency and Formative Utility}
Faculty successfully mitigated initial student apprehension through full transparency, offering an ungraded ``practice mode'' and hosting open discussions about the challenges of writing discriminatory-but-fair exams. The resulting trust in the system was significant enough that the modality transitioned from a purely summative assessment into a highly requested formative tool, with several students proactively requesting access to the AI tester to practice for their traditional, paper-and-pencil final exams.

\subsection{Defending Authorship in the Generative AI Era}
With the proliferation of LLMs capable of producing text that is indistinguishable from undergraduate writing, traditional take-home essays and critical reviews face an existential crisis of validity. One faculty member in the pilot utilized the Socratic Test specifically to neutralize this threat. 

Rather than abandoning a critical review assignment on emerging medical technologies, the instructor deployed the Socratic Test as a mandatory post-submission oral defense. Following the submission of their written critique, students engaged with the AI proctor, which interrogated them on both their specific arguments and the underlying source material. As the instructor noted, this two-step architecture \textit{``ensured that whether or not AI was used to write the paper... they would need to have understood the selected paper and their own critique.''} By shifting the assessment from the production of text to the real-time defense of ideas, the platform restores the validity of long-form writing assignments.

\subsection{Diagnostic Telemetry and the Formative Loop}
In quantitative disciplines, the conversational modality provides instructors with unprecedented diagnostic telemetry. An instructor deploying the platform in a Mechanics of Materials course noted that the system efficiently broke down complex concepts, forcing students to articulate their reasoning rather than relying on ``formula memorization.'' 

Crucially, the conversational format required students to \textit{``explain their process, not just present a final answer.''} This dialogue surfaced specific, cohort-wide conceptual gaps that the instructor was able to dynamically address in the subsequent lecture. Furthermore, this instructor corroborated the Affective Filter hypothesis (Section \ref{sec:affective}), observing that students found the AI interaction \textit{``less intimidating than an oral exam, making the experience feel low-stakes and supportive,''} while still pushing stronger students to deeper levels of explanation.

\section{Conclusion}
The Socratic Test modality transforms the examination from a post-mortem of failure into a dynamic mapping of student capability. By replacing the high-anxiety face-to-face interrogation with a computer-mediated multimodal environment, the platform effectively mitigates the affective filter and neutralizes construct-irrelevant variance. 

Structurally, the architecture resolves the persistent vulnerabilities of AI-mediated assessment. It closes conversational evasion loopholes via continuous mathematical accounting (the Shadow Ledger), absorbs algorithmic hallucination through oversampled evidence buffers, and accommodates diverse pedagogical strategies through dual proctoring modalities. Most critically, by replacing holistic algorithmic evaluation with a deterministic, interaction-level human-AI calibration pipeline, the platform eliminates the ``black box'' of AI grading. By decoupling the prompting framework (Bloom's) from the evaluation framework (SOLO) and applying a non-compensatory mathematical model to Vygotskian scaffolding, educators can deploy highly adaptive, scalable examinations that uphold rigorous academic standards, defend original authorship, and systematically foster a growth mindset.

\bibliographystyle{elsarticle-harv} 
\bibliography{refs}

\newpage
\begin{appendices}

\section{Transcript and Grading Matrix Case Study} \label{app:example}
This appendix demonstrates a simulated Socratic Test interaction utilizing the Dual-Mode architecture running in Stair-Step Mode. The instructor has configured the topic ``Price Controls'' to be worth a maximum of 40 points across three cognitive tiers, with an \textbf{Oversampling Factor of 120\%}. 

Because of the oversampling, the AI's Evidence Buffer requires 12 points of provisional evidence to clear the Foundational tier (Max 10), 18 points to clear the Application tier (Max 15), and 18 points to clear the Synthesis tier (Max 15). The Vertical Gate will not open until the sum of the student's Earned Points and Shadow Ledger Points meets these buffer targets.

\textbf{Dynamic Matrix State for Topic: Price Controls (Max 40 Points)}

\begin{table}[htbp]
    \centering
    \begin{tabular}{l p{2cm} p{2cm} p{2cm} | c}
        \toprule
        \textbf{Exam Stage} & \textbf{Foundational} \newline \textit{(Cap 10, \newline Buffer 12)} & \textbf{Application} \newline \textit{(Cap 15, \newline Buffer 18)} & \textbf{Synthesis} \newline \textit{(Cap 15, \newline Buffer 18)} & \textbf{Gradesheet} \\
        \midrule
        Initial State & 0 Earned, \newline 0 Shadow & 0 Earned, \newline 0 Shadow & 0 Earned, \newline 0 Shadow & \textbf{0 / 40} \\
        \addlinespace
        Int. 1 (Recall) & \textbf{5} Earned, \newline 0 Shadow & 0 Earned, \newline 0 Shadow & 0 Earned, \newline 0 Shadow & \textbf{5 / 40} \\
        \addlinespace
        Int. 2 (Buffer Clear) & \textbf{12} Earned, \newline 0 Shadow & 0 Earned, \newline 0 Shadow & 0 Earned, \newline 0 Shadow & \textbf{10$^\dagger$ / 40} \\
        \addlinespace
        Int. 3 (Student Skip) & 12 Earned, \newline 0 Shadow & 0 Earned, \newline \textbf{8} Shadow & 0 Earned, \newline 0 Shadow & \textbf{10 / 40} \\
        \addlinespace
        Int. 4 (Graceful Exit) & 12 Earned, \newline 0 Shadow & 0 Earned, \newline \textbf{18} Shadow & 0 Earned, \newline 0 Shadow & \textbf{10 / 40} \\
        \addlinespace
        Int. 5 (Hallucination) & 12 Earned, \newline 0 Shadow & 0 Earned, \newline 18 Shadow & 0 Earned, \newline \textbf{12} Shadow & \textbf{10 / 40} \\
        \midrule
        Post-Exam Audit & 12 Earned, \newline 0 Shadow & 0 Earned, \newline 18 Shadow & \textbf{12} Earned, \newline \textbf{0} Shadow & \textbf{22 / 40} \\
        \bottomrule
        \multicolumn{5}{l}{\small \textit{$^\dagger$Points strictly capped at 10 for the final gradebook despite 12 earned in the buffer.}} \\
    \end{tabular}
    \label{tab:appendix_matrix}
\end{table}

\subsection*{Interaction 1: Foundational Recall (Buffer Not Met)}
\textbf{AI Proctor (Bloom 1):} Briefly define a ``price floor'' and give a real-world example. \\
\textbf{Student:} A price floor is a government-mandated minimum price. Minimum wage is an example. \\
\textbf{Real-Time Evaluation:} The student provides a correct definition and an accurate example (\textbf{SOLO 3: Multistructural}). \\
\textbf{Matrix Update:} +5 Points to Foundational Earned. \textit{State Machine Check:} The 5 points fail to satisfy the 12-point Evidence Buffer. The Vertical Gate remains closed.

\subsection*{Interaction 2: Clearing the Oversampled Buffer}
\textbf{AI Proctor (Bloom 2):} Explain how that minimum wage impacts the equilibrium of the labor market. \\
\textbf{Student:} It prevents the wage from dropping to equilibrium, meaning more people want to work than there are jobs, causing unemployment. \\
\textbf{Real-Time Evaluation:} \textbf{SOLO 3 (Multistructural)}. \\
\textbf{Matrix Update:} +7 Points to Foundational Earned. \textit{State Machine Check:} The student now has 12 Earned points. The final grade is strictly capped at 10, but the 12 points exactly meet the 12-point Evidence Buffer. The Vertical Gate opens to the Application tier.

\subsection*{Interaction 3: Evasion and the Shadow Ledger}
\textbf{AI Proctor (Bloom 3):} If the government sets a price floor on agricultural wheat, calculate the resulting deadweight loss assuming a linear supply and demand curve. \\
\textbf{Student:} I don't remember how to calculate that. Can we skip this question? \\
\textbf{AI Action:} The AI complies without penalty and pivots to a new Application prompt. \\
\textbf{Matrix Update (Shadow Ledger):} Skipping a Bloom 3 prompt triggers the Shadow Ledger. According to the baseline matrix (Table \ref{tab:shadow_ledger}), a strong response is SOLO 4. The value $V(3,4)$ is 8 points. 0 Points are added to Earned; +8 Points are added to the Application Shadow Ledger. \textit{State Machine Check:} (0 Earned + 8 Shadow) $<$ 18 Buffer. The Gate remains closed.

\subsection*{Interaction 4: Hint Exhaustion and Graceful Exit}
\textbf{AI Proctor (Bloom 4):} Let's try this conceptually. Explain how the price elasticity of demand for wheat affects the size of the deadweight loss created by the price floor. \\
\textbf{Student:} If demand is highly elastic, people buy less wheat, so the deadweight loss is smaller? \\
\textbf{Scaffolding Triggered:} The AI utilizes the 4-step hierarchy. It offers a General Nudge, a Specific Cue, and a Targeted Scaffold, but the student remains confused. \\
\textbf{Graceful Exit:} The AI provides Direct Instruction: \textit{``Not quite. Elastic demand means consumers are very responsive, leading to a much larger drop in quantity demanded, which increases the deadweight loss. Let's move on.''} \\
\textbf{Matrix Update (Shadow Ledger):} Exhausting the hints results in a discount factor of $\gamma_4 = 0$. The baseline for a Bloom 4 prompt is SOLO 4 (worth 10 points). 0 Points added to Earned; +10 Points added to the Application Shadow Ledger. \textit{State Machine Check:} The Application Buffer now contains 18 points (0 Earned + 18 Shadow). The 18-point Buffer is satisfied. The Vertical Gate opens to the Synthesis tier, locking the student's Application score at 0.

\subsection*{Interaction 5: Hallucination and the Out-of-Scope Protocol}
\textbf{AI Proctor (Bloom 5):} Given this deadweight loss, calculate the exact corresponding rise in national unemployment using the Phillips Curve. \\
\textbf{Student:} The Phillips Curve relates to macroeconomic inflation, not microeconomic deadweight loss. That formula doesn't apply here. \\
\textbf{AI Action:} The AI accepts the out-of-scope flag. To prevent infinite question cycling, the state machine treats this as a Skip and pivots. \\
\textbf{Matrix Update (Shadow Ledger):} According to Table \ref{tab:shadow_ledger}, the baseline value for Bloom 5 is SOLO 4 (worth 12 points). The system logs +12 points to the Synthesis Shadow Ledger. \textit{State Machine Check:} The Synthesis Buffer requires 18 points. With 12 Shadow points added, the AI generates one final replacement Synthesis question to attempt to clear the remaining buffer.

\subsection*{Post-Exam Alignment Pipeline (Step 1 Audit)}
During the batch audit (Section \ref{sec:grading}), the instructor reviews the flagged interaction from Interaction 5. 
\begin{enumerate}
    \item \textbf{Adjudication:} The instructor verifies the AI hallucinated the application of the Phillips Curve and selects ``Student Correct'' on the 4-point scale.
    \item \textbf{Shadow Ledger Conversion:} Recognizing that catching the AI's error proves Extended Abstract mastery (\textbf{SOLO 5}), the grading engine retroactively reverses the 12-point Shadow Ledger penalty logged during the live exam.
    \item \textbf{Mathematical Reward:} Those 12 points are converted directly into the Synthesis \textit{Earned} bucket. The student effectively receives full credit for the botched interaction, ensuring their gradebook accurately reflects their mastery. 
\end{enumerate}

\end{appendices}

\end{document}